\documentclass[11pt]{article}
\usepackage{blois,epsfig}

\begin{document}
\vspace*{4cm}
\title{$C_\ell$ interpolation for cosmological parameter estimation}

\author{ S. BARGOT }

\address{Laboratoire de l'Acc\'el\'erateur Lin\'eaire \\
  IN2P3-CNRS et Universit\'e de Paris-Sud
	BP 34 
	91898 Orsay Cedex}

\maketitle\abstracts{
I will briefly present my work on cosmological parameters estimation.
Classical methods for parameters estimation involve the exploration of the
parameter space on a precalculated grid of cosmological models. Here we try
to estimate the cosmological parameters by using a minimization method associated with the
 interpolation of the 
$C_\ell$ spectrum. We first use a simple multidimensional linear interpolation, and 
show the flaws of this method. We then introduce a new interpolation method, based on 
a physical description of the location of the acoustic peaks in the power spectrum}

As the data available from CMB experiments improve, the estimation of the
cosmological parameters become more and more promising. But 
as the power spectrum estimation increase in precision, we encounter
new problems. The time spent calculating the theoretical power spectrum
and the storage space are two of the main problems. Here we present 
a method based on a grid of pre calculated models with an interpolation method which compute
the power spectrum for any given set of cosmological parameters,
in order to fit the parameters using the most recent data available. 

\section{Cosmological parameters estimation}
Our problem is a classical parameter estimation problem, but in the case 
of the estimation of cosmological parameters from the power spectrum of 
the CMB temperature anisotropies, the difficulty arises from the fact that the 
computation of the cosmological models is a CPU and storage intensive task.
It takes about a minute on a standard PC to compute a single model, so computing 
the statistical quantity to be minimized 
(either the $\chi^2$ or the likelihood function) in the process of 
fitting would be time consuming. One possibility is to calculate the cosmological models 
before the estimation on a discrete grid of the parameter space.  
Methods involving Monte Carlo Markov chains try to reduce the number of
models to be calculated (e.g. Verde ~\cite{2003ApJS..148..195V}) 

In the case of the exploration of the parameter space, we have to compute
the $\chi^2$ or the likelihood function for each point of the grid
in order to obtain the best parameters (e.g. Benoit {\it et al}~\cite{2003A&A...399L..25B}).

In the case of a general minimization algorithm, we have to interpolate the power spectrum in order to
be able to calculate the $\chi^2$ or the likelihood function for any
given set of parameters ${\Omega_i}$. This method allow us to study the 
degeneracies among cosmological parameters via the covariance matrix.

In this analysis, we used a grid with the cosmological parameters presented in 
table \ref{tablecosmo}:
\begin{table}[t]
\caption{\label{tablecosmo}
Cosmological parameters used for the analysis presented in this paper}
\vspace{0.4cm}
\begin{center}
\begin{tabular}{|c|c|c|c|c|c|}
\hline
 & $\Omega_b$ & $\Omega_{CDM}$ & $\Omega_\Lambda$ & $H_0$ & $n_s$ \\
\hline
parameter min & 0.02 & 0.1 & 0. & 50 & 0.9\\
\hline
parameter max & 0.06 & 0.74 & 0.8 & 100 & 1.2\\
\hline
step & 0.002 & 0.02 & 0.01 & 10 & 0.1 \\
\hline
nb of steps & 21 & 31 & 41 & 6 & 4 \\
\hline
\end{tabular}
\end{center}
\end{table}
This has a total of $640584$ nodes and represents 8 Go of disk space.
It was computed using CAMB\footnote{http://camb.info} with $C_\ell$ up to $\ell=3000$. 
 
\section{Cosmological parameters adjustment}
We are trying to fit the cosmological parameters, so instead of calculating
the $\chi^2$ or the likelihood function for each point of the grid 
\ref{tablecosmo}, we want to have a power spectrum for any
given set of parameters ${\Omega_i}$. To realize that, we interpolate
the power spectrum in the parameter space.

We use an ``HyperCube'' to represent and discretize the cosmological 
parameter space. Each point of this HyperCube is defined by the 
cosmological parameters and the corresponding power spectrum. It 
allows us to have an irregular sampling of the parameter space.
\subsection{Simple linear interpolation}
\begin{figure}[h]
\begin{center}
\psfig{figure=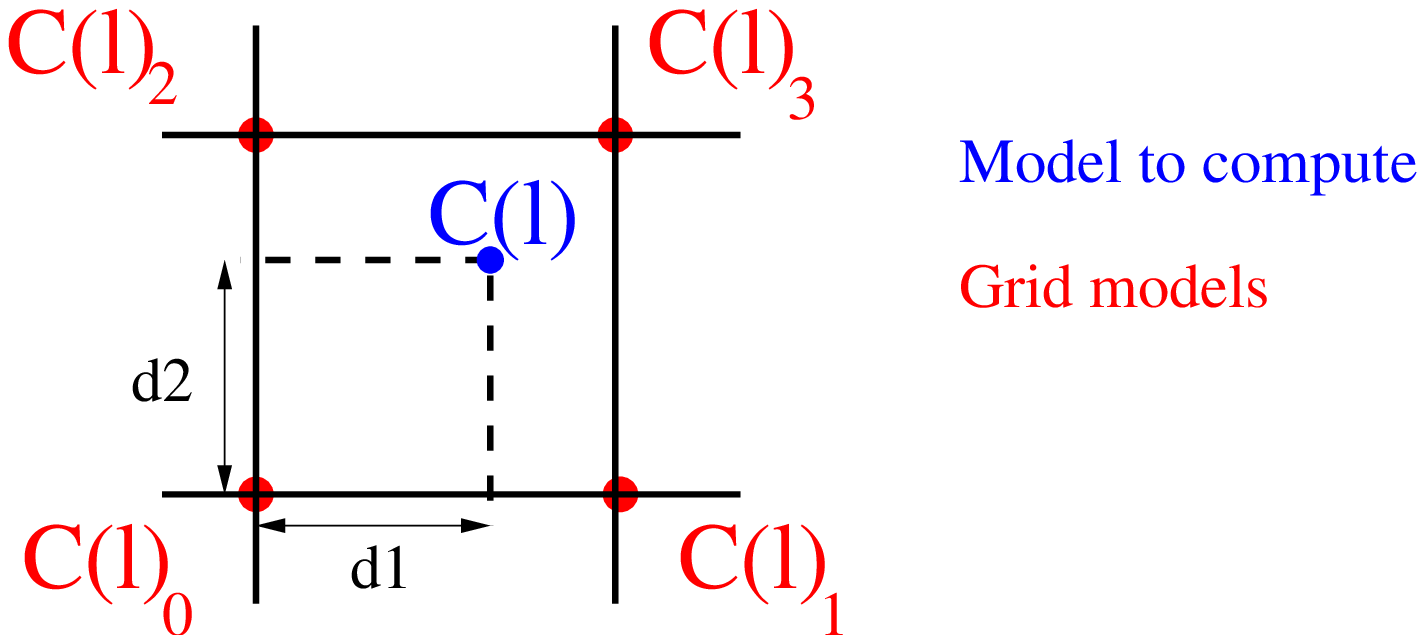,height=1.5in, width=2.7 in}
\psfig{figure=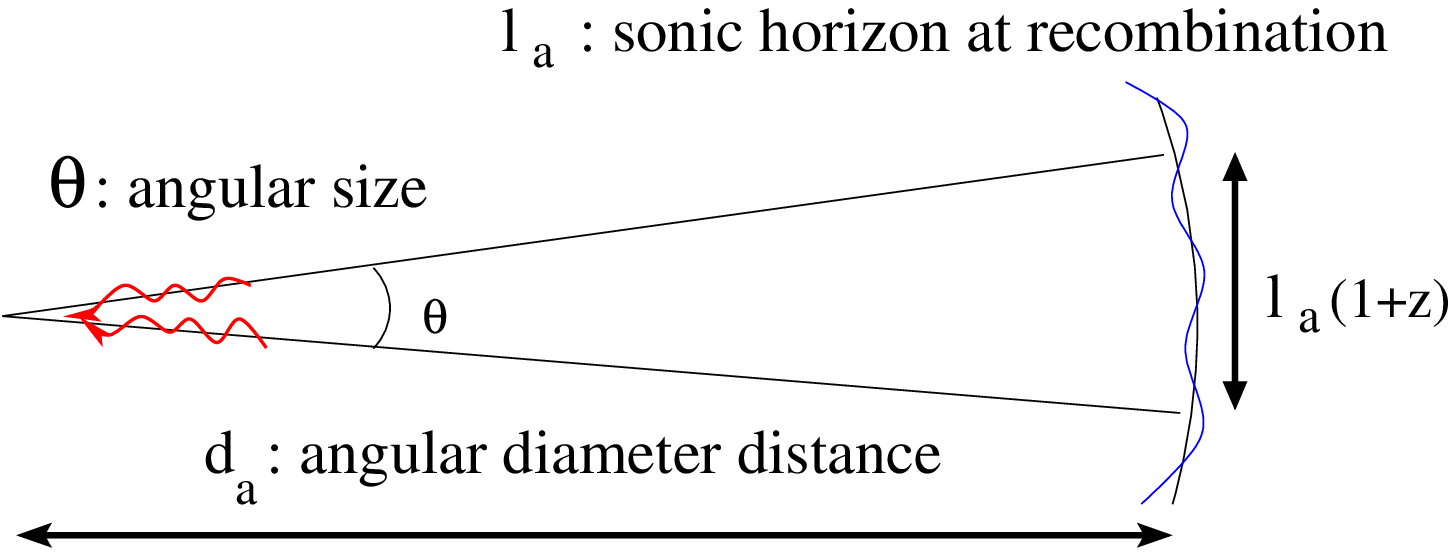,height=1.5in, width=3.5 in}
\caption{a. left : Linear multidimensional interpolation scheme. 
b. right : Acoustic scale.
\label{fig1}}
\end{center}
\end{figure}

We first use a linear multidimensional interpolation. As shown in figure 
\ref{fig1} (left), the power spectrum for a given point ${\Omega_i}$ $C_\ell$ is 
the weighted average of the $C_i(\ell)$ over $p=2^N$ closest neighbours nodes on the grid, where $N$ 
is the parameter hyperspace dimension :
\begin{equation}
\label{eq1}
C_{int}(\ell)= \sum_{i=0}^{2^{N}} w(i) C_i(\ell), \ w(i)=f(d_1,d_2 ... ,d_n).
\end{equation}
In $N$ dimension, the function $w(i)$ is given by the product of $N$ terms, one for each dimension 
which are either $(1-d_i)$ or $(d_i)$. For example, in two dimensions $w(1)=d_1 \times (1-d_2)$ with notations 
according to figure \ref{fig1}.a (left).

\begin{figure}[h]
\begin{center}
\psfig{figure=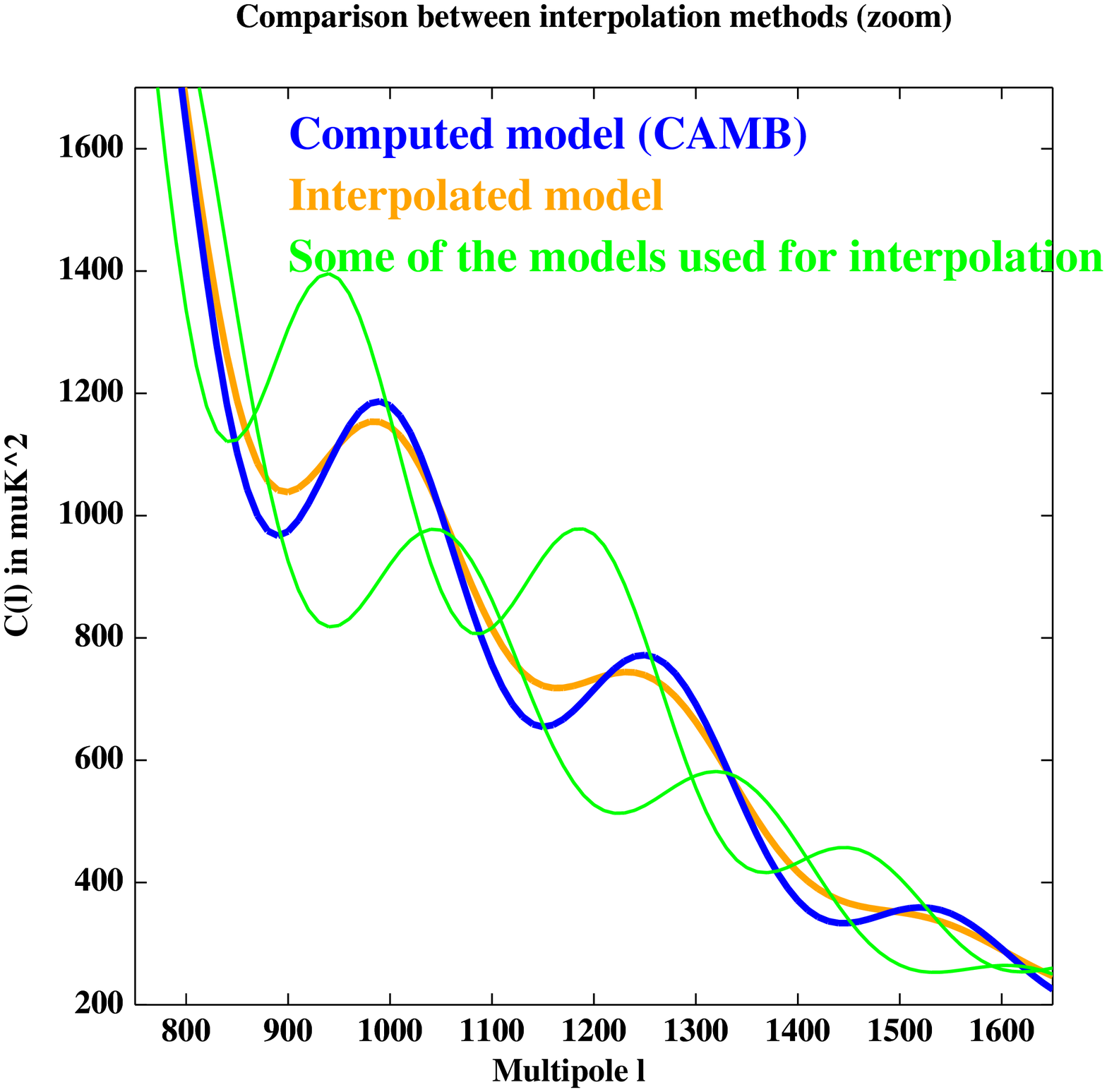,height=2in}
\psfig{figure=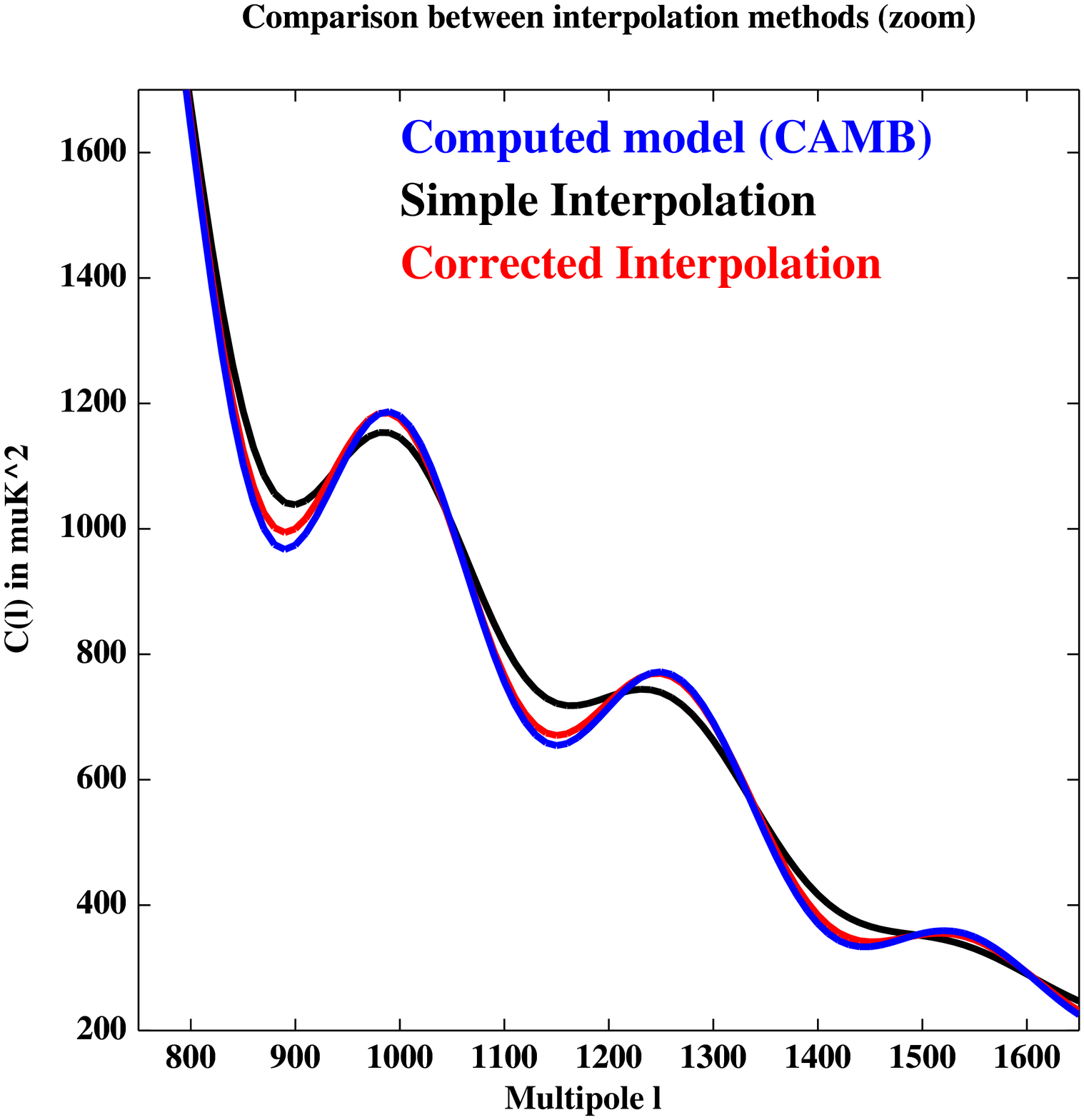,height=2in}
\caption{CAMB computed and interpolated power spectra for $\ell=800$ and $\ell=1600$. 
a. left : Simple multilinear interpolation
b. right : Improved interpolation using the acoustic scale. 
\label{fig3}}
\end{center}
\end{figure}

As it is shown on figure \ref{fig3}.a (left), 
this methods suffers from one major drawback. 
As we use cosmological models with different cosmological parameters, 
we sum power spectra with shifted acoustic peaks. 
The effect is to erase the structure of the peaks, specially at high multipoles. 

\subsection{Acoustic scale}
In order to solve this problem, we use the acoustic scale defined on the figure
\ref{fig1}.

With the notation of the figure, 
the physical size of the acoustic oscillations at decoupling and the angular diameter distance are given by :  
\begin{equation}
l_a=\int_{t_{eq}}^{t_{rec}} c_s dt\ , \ d_a = \frac {1}{\sqrt{\Omega_K}} f\left( \sqrt{\Omega_K}
\int_0^z \frac{dz}{E(z)}\right) , \ \ E(z)=[\Omega_{matter}(1+z)^3+\Omega_{curv}(1+z)^2+\Omega_{\Lambda}]^{1/2}.
\end{equation}
The acoustic scale (in $\ell$-space) is then $as=\pi \frac{d_a}{l_a}$.

\subsection{Improved interpolation}
Our new interpolation scheme is the same as on figure \ref{fig1}.a except that
the  $\ell$ axis is rescaled according to the ratio of the corresponding acoustic scales.
To compute the $C(\ell)$ spectrum for the set of parameter $\{\Omega_t\}$ using the 
$2^N$ grid points , we define $2^N$ rescaling coefficients $\alpha_i(\Omega_t,\Omega_i)=as(\Omega_t)/as(\Omega_i)$
The interpolated $C(\ell)$ spectrum can then be written as :
\begin{equation}
C_{int}(\ell)= \sum_{i=0}^{2^{N}} w(i) C_i(\ell \times \alpha_i, \ \ \ w(i)=f(d_1,d_2 ... ,d_n).
\end{equation}
Figure \ref{fig3}.b shows the improvement of the interpolation result, compared to the simple multilinear
interpolation.

\section{Statistical tests}
In order to quantify the improvement of the interpolation, we have computed two statistical 
quantities :
\begin{itemize}
\item we compute the maximum relative difference $Diff_{max}$ between a CAMB computed CMB spectrum and an 
interpolated one over the $C_\ell$ spectrum 
\item we associate an error $\sigma_\ell$ to each point in the $C(\ell)_{int}$, 
and we compute the $\Delta\chi^2$ between
this ``experiment'' and the CAMB original spectrum 

\begin{equation}
Diff_{max}=Max \left(\frac{C(\ell)_{CAMB}-C_{int}(\ell)}{C(\ell)_{CAMB}}\right)
_\ell ,  
\ \ \Delta \chi^2= \sum_{\ell=\ell_{min}}^{\ell_{max}} \left(\frac{C(\ell)_{CAMB}-C_{int}(\ell)}{\sigma_\ell}\right)^2.
\end{equation}

\end{itemize}

\begin{figure}[h]
\begin{center}
\psfig{figure=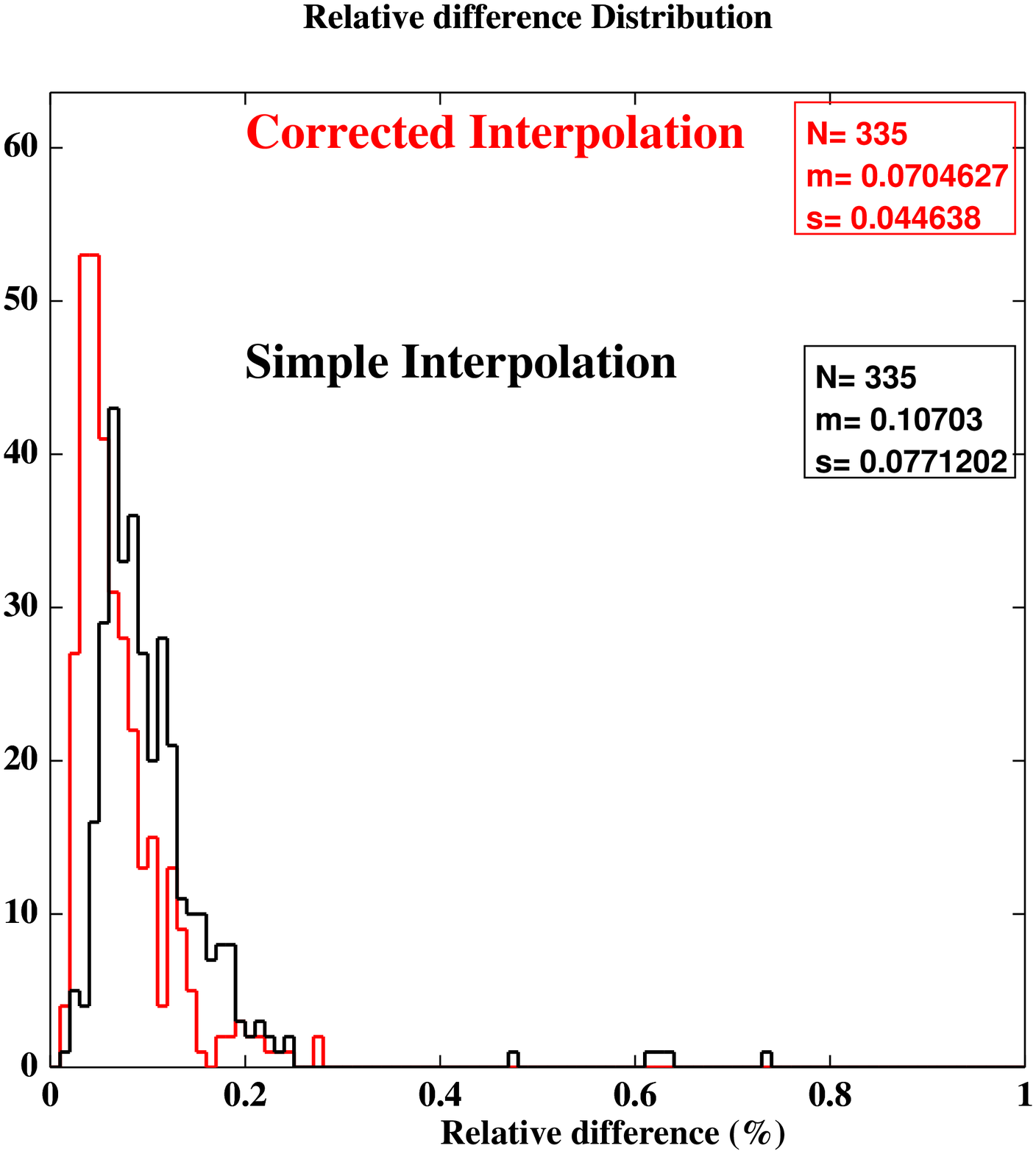,height=2.5in}
\psfig{figure=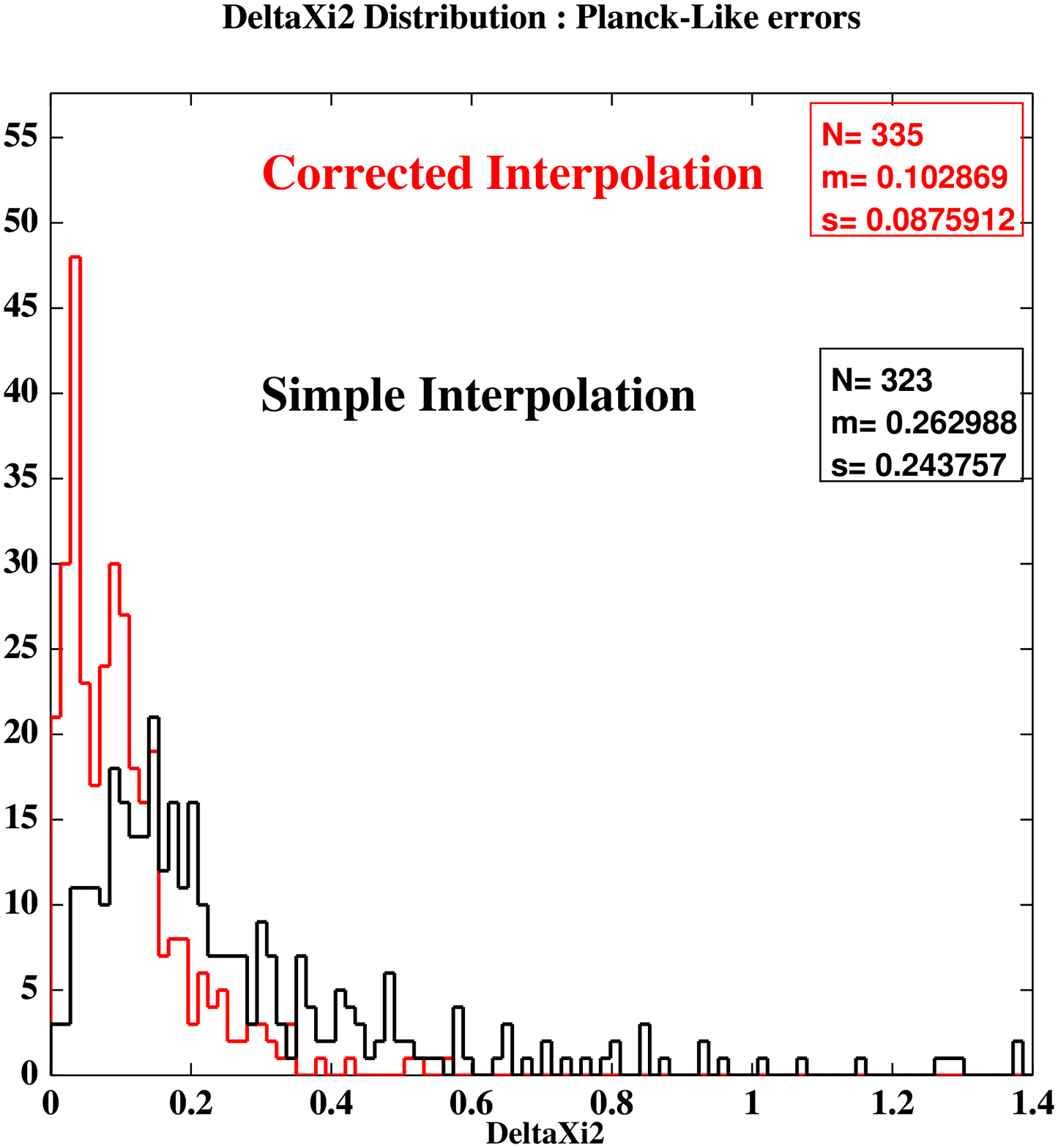,height=2.5in}
\caption{Simple and improved $C(\ell)$ interpolation comparison. 
a. left : relative difference. 
b. right :$\Delta \chi^2$ for Planck-like error bars.
\label{fig6}}
\end{center}
\end{figure}

Figure \ref{fig6} presents the distributions obtained for the relative difference and
for the $\Delta \chi^2$ with error bars similar to the one foreseen for the Planck mission 
(Puget {\it et al}~\cite{planckorangebook}), with 
$\{\Omega_t\}$ randomly distributed over the entire parameter hyperspace.
.

The reduction in the mean value and the spread of the two quantities ($Diff_{max}$ et $\Delta \chi^2$)
obtained by the new interpolation show clearly the improvement.
\section{Conclusion}
We have shown that our improved interpolation method provide a way to diminish the errors
introduced in the cosmological parameters estimation process due to the parameter space 
sampling. Alternatively, for a given precision, it allows to use a smaller grid. It
can be useful in the case of future experiments such as the Planck mission.
Moreover, the parameter fitting is a fast way of estimating cosmological parameter which 
can be used in Monte Carlo simulation of an experiment. The impact of the experimental 
design on the cosmological parameters could then be estimated in this way thanks to the reduction of needed computing 
resources.

\end{document}